# The Fibonacci Sequence Mod $m$
Louis Mello, Ph.D.

**Theorem 1**

Let $F_n$ be the $n^{th}$ number of the Fibonacci series where:
$$F_n = F_{n-1} + F_{n-2} \text{ and } F_0 = o, F_1 = 1 \text{ for } n = 1, 2, 3\ldots \to \infty.$$
We know that $F_n \bmod(p)$ forms a periodic sequence (*vide Theorem 4*). Let $h(p)$ denote the length of the sequence. Let $p$ be a prime such that:
$$p \equiv \{2,3\} \bmod(5)$$
a sufficient and necessary condition to ensure that $h(p)|2p+2$. We shall denote this group $F_1^G$. Let $D = \{d_1, d_2, \ldots, d_k\}$ be the non-empty set of $k$ divisors of $2p+2$. Then for $F_1^G h(p) = \min[d_i]$ such that $F_{d_{i+1}} \equiv 1 \bmod(p)$ and

(a) $d_i \mid \frac{1}{2}p(p+1)$;

(b) $d_i \mid p+1$ and

(c) $d_i \mid 3(p-1)$.

**Theorem 2**

Let $p$ be a prime such that:
$$p \equiv \{1,4\} \bmod(5)$$
a sufficient and necessary condition to ensure that $h(p) | p-1$. We shall denote this group $F_2^G$. If $p$ has a primitive root such that $g^2 \equiv g + 1 \bmod(p)$ (which we shall denote $F_{pr}$) then $h(p) = p-1$. We point out that $g^2 = g + 1 \bmod(p)$ has two roots, notably:
$$g = \frac{1}{2}\sqrt{5} + \frac{1}{2}, g = \frac{1}{2} - \frac{1}{2}\sqrt{5}$$
which both represent the Golden Raio: 1.618033988…and -0.618033988 respectively. If $p$ has no $F_{pr}$ then let $D = \{d_1, d_2, \ldots, d_k\}$ be the non-empty set of $k$ divisors of p-1. Then for $F_2^G$, $h(p) = \min[d_i]$ such that $F_{d_{i+1}} \equiv 1 \bmod(p)$ and

(a) $d_i \mid p+1$ and (b) $d_i \mid \left\lfloor \frac{p}{2} \right\rfloor$ (i.e., $d_i$ is even). If you do not want to previously check for a $F_{pr}$ the results are equivalent.

The section **The 6m Limit** explains why the prime 5 is a special case.

**Theorem 3**

If we replace $p$ (the odd prime) with a Fibonacci number $F_m$ where $m > 3$ we could write $F_n \bmod (F_m)$. The period of the sequence $h(F_m)$ that results from this congruence is given by:

$$\begin{cases} h(F_m) = 2m \Leftrightarrow m \text{ is even} \\ h(F_m) = 4m \Leftrightarrow m \text{ is odd} \end{cases}$$

**Prime Powers:**

If $m = 2^e$ then $h(m) = 3 \times 2^{e-1}$

If $m = 5^e$ then $h(m) = 5^e \times 4$.

And generalizing for prime powers with the exception of 2 and 5:

If $p$ is of the form $2p + 2$ then $h(m) = p^e \dfrac{2p+2}{p}$ (e.g. 3 is of this form).

If $p$ is of the form $p - 1$ then $h(m) = p^e \dfrac{p-1}{p}$

**Why the sequence $F_n \bmod (m)$ is periodic:**

The series $F_n \bmod(m)$ is periodic for two basic reasons:

1. Modulo $m$ there are at most $m^2$ possible pairs of residues, so at some point $\leq m^2$ the residues must repeat;

2. Any pair of consecutive terms of $F_n \bmod(m)$ completely determines the entire sequence.

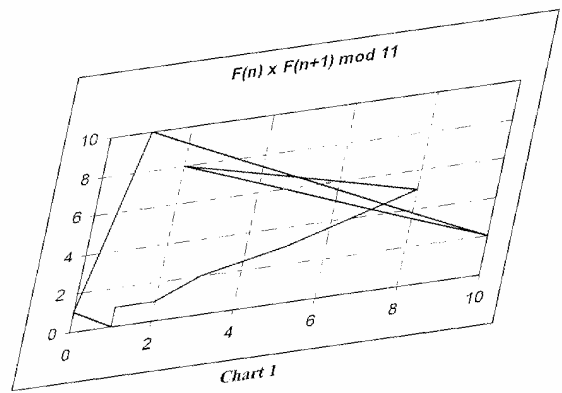

Chart 1

As can be seen from Chart 1 above, by definition of the Fibonacci sequence, for each point on the chart there is only one point that can follow, hence if we revisit any point on the graph the sequence becomes periodic.
Thus we come to the theorem:

**Theorem 4**

$F_n \mod(m)$, where $m \in \mathbb{Z}^+$, forms a periodic sequence. That is, the sequence is periodic and repeats by returning to its initial values.

**Proof**

By definition we have $F_{n+1} - F_n = F_{n-1}$.
So if
$F_{n+1} \circ F_{s+1} \mod(m)$ then $F_{n-1} \circ F_{s-1} \mod(m), \ldots, F_{t-s+1} \circ F_1 \mod(m)$ and $F_{t-s} \circ F_0 \mod(m)$.
Thus, each period must begin with $F_0$. Therefore, we can conclude that $F_n \mod(m)$ forms a simply periodic sequence.

**Generalization for *m***

We also know that, due to Theorem 2, for a non-prime $m > 2$, $h(m)$ can be expressed as follows:

1. For $m > 2$, $h(m)$ is even. Let $h = h(m)$ and consider all equalities to be congruencies modulo $m$. Since $F_h = 0$ and $F_{h+1} = 1$, we can see that $F_{h-1} = 1$. Similarly, $F_{-h+1} = 1$. Now assuming that $h$ is odd, we have, by the following identity: $F_{-t} = (-1)^{t+1} F_t$; that $F_{h-1} = -F_{-(h-1)} = -1$. Since $1 \equiv -1 \mod(m)$ we may conclude that $m = 2$.

2. If $n \mid m$ then $h(n) \mid h(m)$.

3. **Let *m* be factored into primes, $p_1^e, p_2^e, \ldots, p_k^e$. Then** $h(m) = \text{lcm}\left[h\left(p_1^e, \ldots p_k^e\right)\right]$.

    By statement (2) we can observe that $h\left(p_1^e, \ldots, p_k^e\right) \mid h(m)$ for all $p_i^e$ from $i = 1 \ldots k$, so $\text{lcm}\left[h\left(p_{i=1\ldots k}^e\right)\right] \mid h(m)$.

    (a) Let $L = \text{lcm}\left[h\left(p_{i=1\ldots k}^e\right)\right]$. Since $h\left(p_{i=1\ldots k}^e\right) \mid L$ for each $i$ then the sequence $F_n \mod\left(p_{i=1\ldots k}^e\right)$ repeats in blocks of length $L$ (and maybe less) for each $i$. This implies that $F_L = 0$ and $F_{L+1} \equiv 1 \mod\left(p_{i=1\ldots k}^e\right)$ for all $i$.

    (b) Hence, $F_L = 0$ and $F_{L+1} \equiv 1 \mod(m)$ since all $p_{i=1\ldots k}^e$ are co-prime. And, as a result, $h(m) \mid L$.

This last fact tells us that if we can find $h(p^e)$ for primes $p$ then we can find any $m$. That is:
$$h(m) = \text{lcm}\left[h(p_1^e, \ldots, p_k^e)\right].$$

**The 6m Limit:**

Given a positive integer $m$ let $h(m)$ denote the length of the period of the Fibonacci sequence $F_n$ for $n = 1,2,3\ldots$ taken modulo $m$. We wish to prove that $h(m) \leq 6m$ for all $m$ and that equality holds for infinitely many values of $m$.

Let the prime factorization of $m$ be given by:
$$m = \{p_1^e, p_2^e, \ldots, p_k^e\} \tag{0.1}$$

From the period length of a linear recurring sequence mod($m$) it is immediate that:
$$h(m) \text{lcm}\left(h\left(p_j^{e_j}\right)\right) \tag{0.2}$$

which is $\leq \text{lcm}\{p_j^{e_j - 1}, h(p_j)\}$. Thus, any boundary for $h(m)$ must be determined by the periods of the recurrence in the finite fields $Z(p_j)$.

The characteristic polynomial for the Fibonacci and Lucas sequences is given by:
$$q(x) = x^2 - x - 1 \tag{0.3}$$

which splits, in the field $Z(p^2)$ into linear factors $x - a$ and $x - b$. If $a \neq b$ then the $n^{\text{th}}$ element in the sequence has the form:
$$F_n = A \cdot a^n + B \cdot b^n \tag{0.4}$$

for constants $A$ and $B$ determined by the initial values. If $q(x)$ splits in $Z(p)$ then $a,b$ are elements of $Z(p)$ and $h(p) | p - 1$ by FLT. On the other hand, if $q(x)$ is irreducible in $Z(p)$ then the order of the roots of $q(x)$ can be found by noting that:
$$e^p = b \text{ and } -1 = ab = a^{p+1} \tag{0.5}$$

which, of course, implies that $a^{(2(p+1))} = 1$. Thus, $h(p) | 2(p+1)$ for *irreducible primes*. By quadratic reciprocity, $q(x)$ is irreducible over $Z(p)$ if $p = \pm 2 \mod(5)$ and $q(x)$ splits into distinct linear factors over $Z(p)$ if $p = \pm 1 \mod(5)$.

The remaining case is when $q(x)$ has multiple conjugate roots in $Z(p^2)$, which implies that $a = b$ in $Z(p)$. This occurs if and only if $p | |q(x)|$, that means that this will only happen if and only if $p = 5$. In this case the $n^{\text{th}}$ term of the sequence is given by
$$F_n (A + Bn) a^n \tag{0.6}$$

where the constants $A$ and $B$ are again determined by the initial values. Since the periods of $A + B_n$ and $a^n$ divide $p$ and $p - 1$ respectively, the sequence in this case has $h(p) | p(p-1) = 20$ and, indeed, for the Fibonacci sequence we have $h(5) = 20$.

Now, to maximize the value of $\frac{h(m)}{m}$ we must exclude any prime factors $p$ for which $q(x)$ splits into distinct factors in $Z(p)$, since they contribute at best a factor of $\frac{p-1}{p}$.

Therefore we need to consider only products of *irreducible primes* and the special prime 5. If $m$ is a product of only odd *irreducible primes*, then

$$h(m) \le \text{lcm}\left[\left(\frac{p_j+1}{2}\right)p_j^{e_j-1}\right] \le 4m \prod\left(\frac{p_j-1}{2p_j}\right) \tag{0.7}$$

which proves that the ratio is $< 4$ in this case. So, in view of the fact that
$$h(3^n) = 8 \times 3^{n-1} \text{ and } h(2^n) = 3 \times 2^{n-1} \tag{0.8}$$
for both the Fibonacci and Lucas sequences and
$$h(5^n) = \begin{cases} 4 \times 5^n & \text{for the Fibonacci sequence} \\ 4 \times 5^{n-1} & \text{for the Lucas sequence} \end{cases} \tag{0.9}$$

we can easily see that the Fibonacci sequence possesses a maximum value of $\frac{h(m)}{m} = 6$ which occurs $\Leftrightarrow m = 2 \times 5^n$ where $n$ is any positive integer. On the other hand, for the Lucas sequence, there is a maximum value of $\frac{h(m)}{m} = 4$ which occurs $\Leftrightarrow m = 6$.